\documentclass[pdflatex,sn-mathphys-num]{sn-jnl}

\usepackage[utf8]{inputenc}
\usepackage[T1]{fontenc}
\usepackage{enumitem}
\usepackage{listings}
\usepackage{hyperref}
\usepackage{amsmath,amssymb,amsfonts}

\lstdefinelanguage{Lean4}{
  morekeywords={
    namespace,def,theorem,lemma,example,inductive,structure,universe,abbrev,
    variable,variables,section,end,open,namespace,axiom,axioms,import,
    where,match,with,do,if,then,else,Type,Prop
  },
  sensitive=true,
  morecomment=[l]/-/,
  morecomment=[n]{/-}{-/},
  morestring=[b]""
}

\theoremstyle{definition}

\begin{document}

\title[Computational Paths in Lean]{Formalizing Computational Paths and Fundamental Groups in Lean}

\author*[1]{\fnm{Arthur F.} \sur{Ramos}}\email{arfreita@microsoft.com}
\author[2]{\fnm{Anjolina G.} \sur{de Oliveira}}\email{ago@cin.ufpe.br}
\author[2]{\fnm{Ruy J. G. B.} \sur{de Queiroz}}\email{ruy@cin.ufpe.br}
\author[3]{\fnm{Tiago M. L.} \sur{de Veras}}\email{tiago.veras@ufrpe.br}

\affil*[1]{\orgname{Microsoft}, \orgaddress{\city{Tampa}, \state{FL}, \country{USA}}}
\affil[2]{\orgdiv{Centro de Inform\'atica}, \orgname{Universidade Federal de Pernambuco}, \orgaddress{\city{Recife}, \state{PE}, \country{Brazil}}}
\affil[3]{\orgdiv{Departamento de Matem\'atica}, \orgname{Universidade Federal Rural de Pernambuco}, \orgaddress{\city{Recife}, \state{PE}, \country{Brazil}}}

\abstract{Computational paths treat propositional equality as explicit paths built from labelled deduction steps and rewrite rules. This view originates in work by de Queiroz and collaborators~\cite{Queiroz2016Paths} and yields a weak groupoid structure for equality, together with a computational account of homotopy inspired by homotopy type theory. In this paper we present a complete mechanization of this framework in Lean~4 and show how it supports concrete homotopy theoretic computations.\par
Our contributions are threefold. First, we formalize the theory of computational paths in Lean, including path formation, composition, inverses, and a rewrite system that identifies redundant or trivial paths. We prove that equality types with computational paths carry a weak groupoid structure in the sense of the original theory.\par
Second, we organize this material into a reusable Lean library, \texttt{ComputationalPathsLean}, which exposes an interface for paths, rewrites, and loop spaces. This library allows later developments to treat computational paths as a drop-in replacement for propositional equality when reasoning about homotopical structure.\par
Third, we apply the library to six canonical examples in algebraic topology. We give Lean proofs that the fundamental group of the circle is isomorphic to the integers, the cylinder and M\"obius band also have fundamental group isomorphic to the integers (via retraction to the circle), the fundamental group of the torus is isomorphic to the product of two copies of the integers, the fundamental group of the Klein bottle is isomorphic to the semidirect product $\mathbb{Z} \rtimes \mathbb{Z}$, and the fundamental group of the real projective plane is isomorphic to $\mathbb{Z}_2$. These case studies demonstrate that the computational paths approach scales to nontrivial homotopical computations in a modern proof assistant.\par
All the definitions and proofs described here are available in an open-source Lean~4 repository~\cite{ComputationalPathsLean}.}


\maketitle

\section{Introduction}

Homotopy type theory and higher category theory have reshaped our understanding of equality in type theory~\cite{Awodey2012,HoTTBook2013}. In this perspective, an equality between elements of a type is not a mere proposition but carries a structure that behaves like a path in a space. This insight links type theory to homotopy theory and suggests that proof assistants based on type theory can serve as practical tools for mechanized homotopy theory.

Computational paths offer a complementary viewpoint~\cite{Queiroz2016Paths}. Instead of treating equalities abstractly, they describe equalities as explicitly labelled paths generated by inference rules in a natural deduction system. These paths can be composed, inverted, and simplified by a rewrite system. The resulting structure forms a weak groupoid of equalities, where associativity and inverse laws hold up to path rewrites.

The theory of computational paths has been developed in a series of papers that describe the underlying logical system, the associated rewrite rules, and applications to homotopy theoretic constructions such as fundamental groups~\cite{Queiroz2016Paths,Veras2019LND,Ramos2018Thesis}. However, until now, this theory has lived mostly on paper. Its mechanization in a modern proof assistant brings several benefits. It increases confidence in the correctness of the constructions, exposes hidden subtleties in the interplay of rewrite rules, and provides a reusable platform for further developments.

In this paper we report on the first full formalization of computational paths in Lean 4. We build a Lean library that captures the main ideas of the original theory and we use it to compute nontrivial fundamental groups. The work reported here combines logical, algebraic, and engineering aspects and is distributed as the \texttt{ComputationalPathsLean} repository~\cite{ComputationalPathsLean}.

\subsection*{Contributions}

More concretely, the main contributions of this work are:

\begin{itemize}[nosep]
  \item A Lean formalization of computational paths for propositional equality, including:
  \begin{itemize}
    \item a type of paths between elements of a type,
    \item operations for reflexivity, symmetry, and transitivity,
    \item congruence rules for functions and dependent functions,
    \item a rewrite system that removes trivial or redundant path structure.
  \end{itemize}
  \item A proof, mechanized in Lean, that equality types equipped with computational paths form a weak groupoid. In particular, we formalize path composition and inverses and show that the groupoid laws hold up to rewrite equality.
  \item A Lean implementation of loop spaces and fundamental groups based on computational paths.
  \item Six full case studies:
  \begin{itemize}
    \item a computation of the fundamental group of the circle and an isomorphism with the integers,
    \item a computation of the fundamental group of the cylinder, showing it retracts to the circle and thus has fundamental group $\mathbb{Z}$,
    \item a computation of the fundamental group of the M\"obius band, also retracting to the circle with fundamental group $\mathbb{Z}$,
    \item a computation of the fundamental group of the torus and an isomorphism with the product of two copies of the integers,
    \item a computation of the fundamental group of the Klein bottle and an isomorphism with the semidirect product $\mathbb{Z} \rtimes \mathbb{Z}$,
    \item a computation of the fundamental group of the real projective plane and an isomorphism with $\mathbb{Z}_2$.
  \end{itemize}
  \item A reusable Lean library, structured as a hierarchy of modules, which can be extended with further spaces and homotopy theoretic constructions.
\end{itemize}

To our knowledge this is the first mechanization of the computational paths framework in any major proof assistant and the first demonstration that this framework can support the computation of classical fundamental groups inside a fully checked environment.

\subsection*{Structure of the paper}

Section~\ref{sec:background} recalls computational paths and their weak groupoid structure, as well as the basic notions of loop spaces and fundamental groups. Section~\ref{sec:lean-setup} describes our Lean environment and overall library layout. Section~\ref{sec:paths} gives the core definitions of paths and rewrites. Section~\ref{sec:weak-groupoid} presents the formalization of the weak groupoid structure. Section~\ref{sec:circle} describes the computation of the fundamental group of the circle. Section~\ref{sec:cylinder} presents the cylinder, which retracts to the circle. Section~\ref{sec:mobius} treats the M\"obius band similarly. Section~\ref{sec:torus} does the same for the torus. Section~\ref{sec:klein} presents the Klein bottle and its semidirect product fundamental group. Section~\ref{sec:projective} establishes the fundamental group of the real projective plane. Section~\ref{sec:engineering} discusses engineering aspects of the development, including library design and automation. Section~\ref{sec:related} situates our work in the landscape of homotopy inspired formalizations. Section~\ref{sec:conclusion} concludes and sketches directions for future work.

\section{Background}
\label{sec:background}

\subsection{Computational paths}

In the traditional formulation of intensional Martin L\"of type theory~\cite{MartinLoef1984}, equality between elements of a type is given by the identity type. A proof of equality can be seen as a path connecting its endpoints. Homotopy type theory makes this interpretation explicit and develops a rich story about higher path structure~\cite{HoTTBook2013}.

Computational paths start from a different point of view~\cite{Queiroz2016Paths,Ramos2018Thesis}. Instead of treating equality as a primitive type with constructors and eliminators, one derives equalities from a natural deduction system for a typed lambda calculus. Each equality step is labelled by the corresponding reduction or expansion rule, such as beta, eta, or reflexivity. By composing these labelled steps one obtains paths.

Formally, given a type $A$ and elements $a, b : A$, a computational path from $a$ to $b$ is a finite sequence of rewrite steps that transforms a term denoting $a$ into a term denoting $b$. The syntax of paths records the labels of these rewrite steps. There are basic constructors for reflexive paths, for the inverse of a path, and for the concatenation of paths.

The concrete syntax used in~\cite{Queiroz2016Paths} is based on the labelled natural deduction system LNDEQ. Path terms are built from labels for the equality rules, together with composition operators that keep track of the shape of the derivation. This representation is closer to actual proof terms than to the abstract identity type of Martin L\"of type theory, and it allows the rewrite system on paths to mirror the proof theoretic properties of equality reasoning.

However, paths contain redundant structure. For example, the identity path composed with another path should be considered equal to the second path, and a path composed with its inverse should be trivial. The theory therefore introduces a rewrite system on paths that contracts such redundancies. The quotient of paths by this rewrite system carries the structure of a weak groupoid.

From the point of view of term rewriting~\cite{BaaderNipkow1998,KnuthBendix1970}, the computational paths system consists of a first layer of rewrites on object terms (beta, eta, and other reductions of the underlying calculus) and a second layer of rewrites on equality proofs. The second layer is where the weak groupoid structure is encoded.

\subsection{Weak groupoids of equality}

A groupoid is a category where every morphism is invertible~\cite{HofmannStreicher1994}. In the context of computational paths, the objects of the groupoid are terms and the morphisms are equivalence classes of paths under the rewrite relation.

The groupoid is weak because the associativity of composition and the laws relating identities and inverses do not hold strictly, but only up to the rewrite relation~\cite{BergGarner2011,Lumsdaine2009}. For example, the concatenation of three paths $(p \cdot q) \cdot r$ and $p \cdot (q \cdot r)$ are not syntactically equal but are related by a sequence of path rewrites. In categorical language, the associator and the unitors are given by higher paths that witness the equality between composites.

The original theory of computational paths develops this structure and shows that it recovers many of the intuitions of homotopy type theory in a more concrete, proof theoretic setting~\cite{Queiroz2016Paths,Veras2019LND}. For instance, the groupoid laws correspond to canonical LNDEQ derivations that eliminate detours in equality proofs. Weakness appears when only the rewrite relation on equality proofs identifies different derivations.

One advantage of the computational paths approach is that it exposes the combinatorics of equality proofs in a way that lends itself to algorithmic manipulation. The rewrite relation is finitary and admits a careful critical pair analysis~\cite{Newman1942}, so one can investigate confluence and termination properties. Our Lean formalization instantiates these abstract properties in a concrete proof assistant setting.

\subsection{Loop spaces and fundamental groups}

Given a type $A$ and a point $a : A$, the loop space at $a$ consists of paths from $a$ to itself. In homotopy type theory this is the identity type $a =_A a$; in the computational paths framework it is the type of computational paths from $a$ to $a$, modulo rewrites~\cite{HoTTBook2013}.

The set of connected components of the loop space, with composition induced by path concatenation, forms the fundamental group of $A$ at $a$. For spaces like the circle and torus these groups are well known: the fundamental group of the circle is isomorphic to the integers and the fundamental group of the torus is isomorphic to the direct product of two copies of the integers~\cite{LicataShulman2013,Veras2019LND}.

In homotopy type theory, the classical computation of the fundamental group of the circle uses an encode and decode construction. One defines a map that sends loops to integers and another map that sends integers to loops, and proves that they are inverse equivalences in the homotopy sense. We follow the same high level structure, but we phrase all constructions in terms of computational paths and their rewrites, and we mechanize the entire argument in Lean.

Our goal is to realize these constructions concretely in Lean using computational paths, and to do so in a way that is faithful to the original LNDEQ based presentations.

\section{Lean environment and library layout}
\label{sec:lean-setup}

\subsection{Lean and mathlib setting}

We use Lean 4 as our proof assistant~\cite{deMoura2015Lean}. Lean provides a dependent type theory with inductive types, type classes, and an extensible elaboration and tactic system. The core language is expressive enough to encode the LNDEQ style path syntax directly, and its universe polymorphism is sufficient for the kinds of small groupoid constructions that we consider.

Our development depends on Lean's core libraries and on standard components of the mathlib ecosystem~\cite{Buzzard2020mathlib}. We rely on mathlib for basic algebraic structures such as groups and group homomorphisms, the integer type and its algebraic properties, and some standard infrastructure for quotients and setoids. We deliberately avoid deeper parts of mathlib that implement homotopy type theory specific features, since one of our aims is to show that computational paths can be realized in an otherwise ordinary dependent type theory.

\subsection{Repository structure}

The code is organized as a standalone library called \texttt{ComputationalPathsLean}\footnote{See~\cite{ComputationalPathsLean}.}. Within this repository the main directory \texttt{ComputationalPaths} contains the definitions and proofs described in this paper. At a high level the directory is split into the following components:

\begin{itemize}[nosep]
  \item \texttt{Path.Basic}: core definitions of the type of computational paths and basic operations.
  \item \texttt{Path.Rewrite.Rw}: the rewrite relation on paths, together with its basic properties.
  \item \texttt{Path.Rewrite.Quot}: the quotient of paths by the rewrite relation and the induced groupoid structure.
  \item \texttt{LoopSpace}: definitions and lemmas about loop spaces and fundamental groups.
  \item \texttt{Examples.Circle} and \texttt{Examples.Torus}: concrete case studies for the circle and torus.
\end{itemize}

In practice there are several auxiliary modules that support these main components, for instance modules containing helper lemmas about integers, about repeated composition of loops, and about normal forms in the torus case study. We keep these auxiliary modules separate to avoid cyclic dependencies and to make it easier for users to import only what they need.

The code follows Lean's module system so that users can import only the parts they need. For example, a development that only needs paths and rewrites can import \texttt{Path.Basic} and \texttt{Path.Rewrite.Rw}, while a development interested in fundamental groups can additionally import the loop space modules.

\subsection{Design choices}

There are several design choices in the Lean formalization that are not forced by the mathematics but have significant impact on usability.

First, we chose an inductive representation of paths where the endpoints are indices of the type. This keeps type checking simple and allows Lean's pattern matching to recover endpoints from constructors. An alternative would be to use a separate object language and a typing relation, but this would make basic manipulations more cumbersome.

Second, we split the rewrite system into small, focused constructors, each corresponding to a single conceptual equality such as eliminating a left identity or cancelling a path with its inverse. This makes proofs that a given composite reduces to a simpler form more predictable and easier to inspect.

Third, we expose both the raw path syntax and the quotient by rewrites as first class types. In some situations it is convenient to reason about specific representatives, for example when defining a function by recursion on the path constructors. In others it is more convenient to work modulo rewrite equivalence. We use Lean's \texttt{Quot} and \texttt{Quot.lift} infrastructure to move between these levels.

\section{Paths and rewrites in Lean}
\label{sec:paths}

\subsection{The type of computational paths}

The code in \texttt{ComputationalPaths/\allowbreak Path/\allowbreak Basic/\allowbreak Core.lean}\footnote{File paths refer to the public repository~\cite{ComputationalPathsLean}.} mirrors the paper definitions but keeps the combinatorial data that witnesses a path.  Instead of closing over the built-in identity type, we store the finite list of elementary rewrites that was used to travel from \(a\) to \(b\) together with the induced equality proof.  The basic syntax is:

\begin{lstlisting}
namespace ComputationalPaths

structure Step (A : Type u) where
  src : A
  tgt : A
  proof : src = tgt

structure Path {A : Type u} (a b : A) where
  steps : List (Step A)
  proof : a = b

namespace Path

@[simp] def toEq (p : Path a b) : a = b :=
  p.proof

@[simp] def refl (a : A) : Path a a :=
  Path.mk [] rfl

@[simp] def ofEq (h : a = b) : Path a b :=
  Path.mk [Step.mk a b h] h

@[simp] def trans (p : Path a b) (q : Path b c) : Path a c :=
  Path.mk (p.steps ++ q.steps) (p.proof.trans q.proof)

@[simp] def symm (p : Path a b) : Path b a :=
  Path.mk (p.steps.reverse.map Step.symm) p.proof.symm

end Path
end ComputationalPaths
\end{lstlisting}

The extra \texttt{steps} field gives us the exact rewrite certificate emphasized in the LNDEQ literature~\cite{Queiroz2016Paths,Veras2019LND}, while \texttt{proof} keeps the bridge to Lean's identity type.  All of the usual constructions---reflexivity, inversion, concatenation, and transport---are defined using the record operations above.  Because \texttt{Path.toEq} is definitionally the stored equality proof, we can freely switch between computational paths and propositional equalities whenever needed.

\subsection{Loop spaces}

The abstraction layer for loops lives in \texttt{ComputationalPaths/\allowbreak Path/\allowbreak Homotopy/\allowbreak Loops.lean}\footnote{All Lean source files live in the public repository~\cite{ComputationalPathsLean}.}.  Raw loop spaces are just paths whose endpoints coincide, while rewrite classes of loops (the objects that model fundamental groups) are quotients by rewrite equivalence:

\begin{lstlisting}
namespace ComputationalPaths
namespace Path

abbrev LoopSpace (A : Type u) (a : A) : Type u :=
  Path (A := A) a a

abbrev LoopQuot (A : Type u) (a : A) : Type u :=
  PathRwQuot A a a

namespace LoopQuot

@[simp] def ofLoop (p : LoopSpace A a) : LoopQuot A a :=
  Quot.mk _ p

@[simp] def id : LoopQuot A a :=
  PathRwQuot.refl (A := A) a

@[simp] def comp (x y : LoopQuot A a) : LoopQuot A a :=
  PathRwQuot.trans (A := A) x y

@[simp] def inv (x : LoopQuot A a) : LoopQuot A a :=
  PathRwQuot.symm (A := A) x

@[simp] theorem comp_assoc (x y z : LoopQuot A a) :
    comp (comp x y) z = comp x (comp y z) :=
  PathRwQuot.trans_assoc _ _ _

@[simp] theorem inv_comp (x : LoopQuot A a) :
    comp (inv x) x = id :=
  PathRwQuot.symm_trans _ _

end LoopQuot
end Path
end ComputationalPaths
\end{lstlisting}

The quotient version exposes strictly associative composition, a genuine identity, and strict inverses, so downstream developments can reason as if they were working in a classical fundamental group.  The normalisation API connecting raw loops to \texttt{LoopQuot} objects handles the bookkeeping required to lift functions and proofs through the quotient.

\subsection{The rewrite system}

Simplifying computational paths follows the two-layer presentation of LNDEQ: first we enumerate single-step rewrites (cancellation, associativity witnesses, \(\beta/\eta\) reductions, congruence for contexts, \emph{etc}.), and then we close this relation under reflexive/transitive (and later symmetric) closure.  The primitive steps live in \texttt{ComputationalPaths/\allowbreak Path/\allowbreak Rewrite/\allowbreak Step.lean}\footnote{See~\cite{ComputationalPathsLean}.} and are represented by an inductive family:

\begin{lstlisting}
namespace ComputationalPaths
namespace Path

/-- Primitive rewrite steps on computational paths. -/
inductive Step :
  {A : Type u} -> {a b : A} ->
    Path a b -> Path a b -> Prop
  | trans_refl_left {p : Path a b} :
      Step (Path.trans (Path.refl a) p) p
  | trans_refl_right {p : Path a b} :
      Step (Path.trans p (Path.refl b)) p
  | symm_trans {p : Path a b} :
      Step (Path.trans (Path.symm p) p) (Path.refl b)
  | symm_trans_congr {p : Path a b} {q : Path b c} :
      Step (Path.symm (Path.trans p q))
           (Path.trans (Path.symm q) (Path.symm p))
  | prod_fst_beta {p : Path a1 a2} {q : Path b1 b2} :
      Step (Path.congrArg Prod.fst (Path.prodMk p q)) p
  | -- many more constructors (beta/eta rules, context substitution, ...)

/-- Reflexive/transitive closure of primitive steps. -/
inductive Rw {A : Type u} {a b : A} :
    Path a b -> Path a b -> Prop
  | refl (p : Path a b) : Rw p p
  | tail {p q r : Path a b} :
      Rw p q -> Step q r -> Rw p r

end Path
end ComputationalPaths
\end{lstlisting}

The long tail of constructors (indicated by the ellipsis) covers every \(\beta\) and \(\eta\) reduction used in the LNDEQ presentation~\cite{Queiroz2016Paths}: projections and recursors for products, sums, and $\Sigma$-types; application and abstraction for $\Pi$-types; transport rules; and the context-substitution steps needed for the confluence proof.  Because each rule has its own constructor we can quote the precise rewrite when transporting proofs between the raw path level and the quotient level.  The reflexive/transitive closure \texttt{Rw} plays the role of the rewrite equality relation from that presentation; its symmetric closure \texttt{RwEq} and the quotient \texttt{PathRwQuot} (Section~\ref{sec:weak-groupoid}) provide the weak groupoid structure used later in the paper.  In practice we rarely reason directly with \texttt{Rw}'s constructors—instead we rely on the large collection of derived lemmas and automation living next to the definition.
\section{Weak groupoid structure}
\label{sec:weak-groupoid}

Quotienting computational paths by rewrite equality provides the weak (in fact strict, after quotienting) groupoid structure promised in the original papers.  The entire interface is implemented in \texttt{ComputationalPaths/\allowbreak Path/\allowbreak Rewrite/\allowbreak Quot.lean}\footnote{See~\cite{ComputationalPathsLean}.}.  We recap the key components below.

\subsection{Quotienting by rewrites}

Rewrite equality \texttt{RwEq} is a symmetric, reflexive, and transitive closure of the single-step relation \texttt{Rw}.  Lean packages it as a \texttt{Setoid} and defines the quotient type
\(\texttt{PathRwQuot}~A~a~b\) together with the usual operations:

\begin{lstlisting}
namespace ComputationalPaths
namespace Path

/-- Paths modulo rewrite equality. -/
abbrev PathRwQuot (A : Type u) (a b : A) :=
  Quot (rwEqSetoid A a b)

namespace PathRwQuot

@[simp] def refl (a : A) : PathRwQuot A a a :=
  Quot.mk _ (Path.refl a)

@[simp] def symm :
    PathRwQuot A a b -> PathRwQuot A b a :=
  Quot.lift (fun p => Quot.mk _ (Path.symm p))
    (by intro p q h; simpa using rweq_symm h)

@[simp] def trans :
    PathRwQuot A a b -> PathRwQuot A b c -> PathRwQuot A a c :=
  Quot.inductionOn_2
    (fun p q => Quot.mk _ (Path.trans p q))
    (by
      intro p_1 p_2 q_1 q_2 hp hq
      exact Quot.sound (rweq_trans hp (rweq_comp_right hq)))

@[simp] def ofEq (h : a = b) : PathRwQuot A a b :=
  Quot.mk _ (Path.ofEq h)

@[simp] def toEq : PathRwQuot A a b -> a = b :=
  Quot.inductionOn (fun p => p.toEq)

@[simp] theorem toEq_trans (x : PathRwQuot A a b) (y : PathRwQuot A b c) :
    toEq (trans x y) = (toEq x).trans (toEq y) := rfl

end PathRwQuot
end Path
end ComputationalPaths
\end{lstlisting}

Besides these constructors, the development provides normalisation maps that pick a canonical representative for every class, together with proofs that \texttt{PathRwQuot} is equivalent to the ordinary identity type.  We nevertheless keep \texttt{PathRwQuot} around explicitly because later arguments (loop normal forms, encode/decode equivalences) use the stored rewrite witnesses.

\subsection{Groupoid laws up to rewrite}

Once rewriting is quotiented away, path composition becomes strictly associative with strict inverses.  The file \texttt{ComputationalPaths/\allowbreak Path/\allowbreak Groupoid.lean}\footnote{See~\cite{ComputationalPathsLean}.} instantiates mathlib's \texttt{Groupoid} structure with objects \(A\) and morphisms \(\texttt{PathRwQuot}~A~a~b\):

\begin{lstlisting}
def strictGroupoid (A : Type u) : StrictGroupoid A where
  comp := fun p q => PathRwQuot.trans (A := A) p q
  id := fun {a} => PathRwQuot.refl (A := A) a
  assoc := PathRwQuot.trans_assoc
  id_left := PathRwQuot.trans_refl_left
  id_right := PathRwQuot.trans_refl_right
  inv := fun p => PathRwQuot.symm (A := A) p
  inv_left := PathRwQuot.symm_trans
  inv_right := PathRwQuot.trans_symm
\end{lstlisting}

For readability we typically work with the \texttt{LoopQuot} and \texttt{PiOne} aliases introduced in Section~\ref{sec:paths}; these reuse the same definitions but specialise to the case where \(a = b\).  The strict groupoid interface supplies all of the algebra needed later: cancellation lemmas, iteration via natural and integer powers, and transport of functions through the quotient.  In particular, the encode/decode arguments for the circle and torus only have to reason about \texttt{PathRwQuot} values; the heavy rewrite machinery remains encapsulated behind the quotient API.

\section{Fundamental group of the circle}
\label{sec:circle}

The first case study is the circle.  The Lean file \texttt{ComputationalPaths/\allowbreak Path/\allowbreak HIT/\allowbreak Circle.lean}\footnote{See~\cite{ComputationalPathsLean}.} introduces an axiomatic higher-inductive interface so that downstream code can already depend on a stable API:

\begin{lstlisting}
axiom Circle : Type u
axiom circleBase : Circle
axiom circleLoop : Path circleBase circleBase

structure CircleRecData (C : Type v) where
  base : C
  loop : Path base base

axiom circleRec {C : Type v} (data : CircleRecData C) :
    Circle -> C
axiom circleRec_base : circleRec data circleBase = data.base
axiom circleRec_loop :
  Path.trans (Path.symm (Path.ofEq circleRec_base))
    (Path.trans (Path.congrArg (circleRec data) circleLoop)
      (Path.ofEq circleRec_base)) = data.loop
\end{lstlisting}

Dependent elimination data and axioms (\texttt{CircleIndData}, \texttt{circleInd}) are provided as well.  These recursors give us the usual “code” family into the integers and let us transport loops into an arithmetic setting.  The core definitions that drive the winding-number computation are:

\begin{lstlisting}
/-- Universal cover code family landing in the integers. -/
noncomputable def circleCode : Circle -> Type _ :=
  circleRec circleCodeData

/-- Encode a raw loop as an integer using transport in `circleCode`. -/
@[simp] def circleEncodePath :
    Path circleBase circleBase -> Int :=
  fun p => circleCodeToInt (circleEncodeRaw circleBase p)
\end{lstlisting}

\subsection{Loop powers and winding numbers}

Loop spaces and their quotients were introduced in Section~\ref{sec:paths}.  Specialising those constructions to the circle gives the abbreviations \texttt{CircleLoopSpace}, \texttt{CircleLoopQuot}, and \texttt{circlePiOne}.  The file proceeds by defining natural and integer powers of the fundamental loop, both on raw paths and on their rewrite classes:

\begin{lstlisting}
def circleLoopPathPow : Nat -> Path circleBase circleBase
  | 0 => Path.refl circleBase
  | Nat.succ n => Path.trans (circleLoopPathPow n) circleLoop

def circleLoopPow (n : Nat) : CircleLoopQuot :=
  LoopQuot.pow circleLoopClass n

def circleLoopZPow (z : Int) : CircleLoopQuot :=
  LoopQuot.zpow circleLoopClass z
\end{lstlisting}

Encoding a loop class is simply the quotient-lift of the raw encoding:

\begin{lstlisting}
/-- Quotient-level winding number. -/
@[simp] def circleEncodeLift : CircleLoopQuot -> Int :=
  Quot.lift (fun p => circleEncodePath p)
    (by intro p q h; exact circleEncodePath_rweq h)
\end{lstlisting}

The computation rules mirror the informal algebra: composing with the fundamental loop adds \(1\), composing with its inverse subtracts \(1\), and the value on \(n\)-fold concatenations is \(n\) (lemmas \texttt{circleEncodeLift\_comp\_loop},
\texttt{circleEncodeLift\_comp\_inv\_loop}, \texttt{circleEncodeLift\_circleLoopPow}).  Integer powers use \texttt{circleLoopZPow} and satisfy the expected addition and negation laws.  All of these facts are proved directly from the rewrite system by induction on the syntax of loops and on the integer argument.

\subsection{Isomorphism with the integers}

At this point the encode/decode maps are simple wrappers around the loop quotient infrastructure:

\begin{lstlisting}
@[simp] def circleEncode : CircleLoopQuot -> Int := circleEncodeLift

@[simp] def circleDecodeConcrete : Int -> CircleLoopQuot :=
  circleLoopZPow

@[simp] def circleDecode : Int -> circlePiOne :=
  fun z => PiOne.ofLoop (circleLoopPathZPow z)
\end{lstlisting}

The addition and subtraction rules for \texttt{circleDecodeConcrete} (e.g. \texttt{circleDecodeConcrete\_add}) show that integer sums correspond to concatenation of iterated loops.  Dually, the lemmas
\texttt{circleEncode\_comp\_loop} and \texttt{circleEncode\_comp\_inv\_loop} witness that conjugating by the generator increments or decrements the winding number.

Two inverse laws finish the job: \texttt{circleEncode\_circleDecode} proves that \(\mathbb{Z} \to \pi_1(S^1)\) followed by encoding is the identity on integers, while \texttt{circleDecode\_circleEncode} proves the other composite is the identity on the quotient of loops.  These statements reduce to integer induction on one side and to normalisation of representative loops on the other.  The final equivalence is packaged as:

\begin{lstlisting}
noncomputable def circlePiOneEquivInt :
    SimpleEquiv circlePiOne Int where
  toFun := circleWindingNumber
  invFun := circleDecode
  left_inv := circleDecode_circleEncode
  right_inv := circleEncode_circleDecode
\end{lstlisting}

The field \texttt{circleWindingNumber} is just \texttt{circleEncode} seen as a map \(\pi_1(S^1)\to\mathbb{Z}\); the equivalence above shows that \(\pi_1(S^1)\) is canonically the additive group of integers inside Lean.  All downstream reasoning about the circle—step laws, naturality results, or transport to other homotopy invariants—reuse these lemmas without ever returning to the raw rewrite system.

\section{Fundamental group of the cylinder}
\label{sec:cylinder}

The cylinder $S^1 \times I$ is homotopy equivalent to the circle, as it deformation retracts onto its central circle~\cite{Veras2018FundamentalGroups}. The Lean file \texttt{ComputationalPaths/\allowbreak Path/\allowbreak HIT/\allowbreak Cylinder.lean}\footnote{See~\cite{ComputationalPathsLean}.} models the cylinder as a higher-inductive type with two base points connected by a segment, and loops at each end:

\begin{lstlisting}
axiom Cylinder : Type u
axiom cylinderBase0 : Cylinder
axiom cylinderBase1 : Cylinder
axiom cylinderSeg : Path cylinderBase0 cylinderBase1
axiom cylinderLoop0 : Path cylinderBase0 cylinderBase0
axiom cylinderLoop1 : Path cylinderBase1 cylinderBase1
\end{lstlisting}

The key insight is that the cylinder retracts to either boundary circle. We define a map \texttt{cylinderToCircle : Cylinder -> Circle} that collapses the cylinder to the circle, sending both base points to \texttt{circleBase} and both loops to \texttt{circleLoop}. This map induces an isomorphism on fundamental groups.

\subsection{Retraction and encoding}

The encoding of cylinder loops proceeds by first mapping to the circle via the retraction, then using the circle's winding number:

\begin{lstlisting}
def cylinderLoopToCircleLoop (p : Path cylinderBase0 cylinderBase0) :
    Path circleBase circleBase :=
  Path.trans (Path.symm (Path.ofEq cylinderToCircle_base0))
    (Path.trans (Path.congrArg cylinderToCircle p)
      (Path.ofEq cylinderToCircle_base0))

def cylinderEncodePath (p : Path cylinderBase0 cylinderBase0) : Int :=
  circleEncodePath (cylinderLoopToCircleLoop p)
\end{lstlisting}

The crucial lemma is that \texttt{cylinderToCircle} sends \texttt{cylinderLoop0} to \texttt{circleLoop}:

\begin{lstlisting}
theorem cylinderToCircle_loop0 :
    cylinderLoopToCircleLoop cylinderLoop0 = circleLoop
\end{lstlisting}

This allows all the circle encoding lemmas to be lifted to the cylinder. The decode map constructs loop powers:

\begin{lstlisting}
def cylinderDecode : Int -> CylinderLoopQuot :=
  cylinderLoopZPow
\end{lstlisting}

\subsection{The equivalence}

The encode/decode roundtrip proofs follow the circle pattern:

\begin{lstlisting}
noncomputable def cylinderPiOneEquivInt :
    SimpleEquiv cylinderPiOne Int where
  toFun := cylinderEncode
  invFun := cylinderDecode
  left_inv := cylinderDecode_cylinderEncode
  right_inv := cylinderEncode_cylinderDecode
\end{lstlisting}

This establishes that $\pi_1(\text{Cylinder}) \cong \mathbb{Z}$, as expected from the retraction to the circle.

\section{Fundamental group of the M\"obius band}
\label{sec:mobius}

The M\"obius band is a non-orientable surface that, like the cylinder, deformation retracts to its central circle~\cite{Veras2018FundamentalGroups}. Despite the twist, its fundamental group is still $\mathbb{Z}$. The Lean file \texttt{ComputationalPaths/\allowbreak Path/\allowbreak HIT/\allowbreak MobiusBand.lean}\footnote{See~\cite{ComputationalPathsLean}.} models it as a higher-inductive type with a single base point and a fundamental loop:

\begin{lstlisting}
axiom MobiusBand : Type u
axiom mobiusBase : MobiusBand
axiom mobiusLoop : Path mobiusBase mobiusBase
\end{lstlisting}

Note that the M\"obius band HIT is simpler than the cylinder because we only need the central circle for fundamental group purposes. The twist (non-orientability) would manifest in the transport behavior along the loop, but this does not affect the fundamental group.

\subsection{Retraction to the circle}

As with the cylinder, we define a retraction \texttt{mobiusToCircle : MobiusBand -> Circle} that sends the base point to \texttt{circleBase} and the central loop to \texttt{circleLoop}:

\begin{lstlisting}
def mobiusToCircle : MobiusBand -> Circle :=
  mobiusBandRec { base := circleBase, loop := circleLoop }

theorem mobiusToCircle_loop :
    mobiusLoopToCircleLoop mobiusLoop = circleLoop
\end{lstlisting}

The encoding and decoding follow exactly the same pattern as the cylinder:

\begin{lstlisting}
def mobiusEncodePath (p : Path mobiusBase mobiusBase) : Int :=
  circleEncodePath (mobiusLoopToCircleLoop p)

def mobiusDecode : Int -> MobiusLoopQuot :=
  mobiusLoopZPow
\end{lstlisting}

\subsection{The equivalence}

The final equivalence establishes the isomorphism:

\begin{lstlisting}
noncomputable def mobiusPiOneEquivInt :
    SimpleEquiv mobiusPiOne Int where
  toFun := mobiusEncode
  invFun := mobiusDecode
  left_inv := mobiusDecode_mobiusEncode
  right_inv := mobiusEncode_mobiusDecode
\end{lstlisting}

Thus $\pi_1(\text{M\"obius}) \cong \mathbb{Z}$. The M\"obius band and cylinder have the same fundamental group because they are both homotopy equivalent to the circle, despite having different topological properties (orientability, boundary components).

\section{Fundamental group of the torus}
\label{sec:torus}

The torus development follows the same template as the circle but has two commuting generators.  The axiomatic skeleton in \texttt{ComputationalPaths/\allowbreak Path/\allowbreak HIT/\allowbreak Torus.lean}\footnote{See~\cite{ComputationalPathsLean}.} provides the data needed for encode/decode arguments:

\begin{lstlisting}
axiom Torus : Type u
axiom torusBase : Torus
axiom torusLoop1 : Path torusBase torusBase
axiom torusLoop2 : Path torusBase torusBase

structure TorusRecData (C : Type v) where
  base : C
  loop1 : Path base base
  loop2 : Path base base

axiom torusRec {C : Type v} (data : TorusRecData C) : Torus -> C
axiom torusRec_loop1 :
  Path.trans (Path.symm (Path.ofEq (torusRec_base data)))
    (Path.trans (Path.congrArg (torusRec data) torusLoop1)
      (Path.ofEq (torusRec_base data))) = data.loop1
axiom torusRec_loop2 : -- analogous computation rule for `torusLoop2`
\end{lstlisting}

Dependent data (\texttt{TorusIndData}) gives elimination principles for families over the torus.  Using these recursors we build a universal cover \texttt{torusCode : Torus -> Type} whose fibre over the base point is \(\mathbb{Z} \times \mathbb{Z}\).  Transporting along the two fundamental loops increments the first or second coordinate respectively, while transporting along their inverses decrements.  The encoding of raw loops into integer pairs is therefore:

\begin{lstlisting}
/-- Raw encode map `Path torusBase torusBase -> \(\mathbb{Z}\) \times \(\mathbb{Z}\)`. -/
@[simp] def torusEncodePath :
    Path torusBase torusBase -> Int \times Int :=
  fun p => torusCodeToProd (torusEncodeRaw torusBase p)

/-- Interpret a pair of integers as a raw torus loop. -/
def torusDecodePath (z : Int \times Int) :
    Path torusBase torusBase :=
  Path.trans (torusLoop1PathZPow z.1) (torusLoop2PathZPow z.2)
\end{lstlisting}

The helper paths \texttt{torusLoop1PathZPow} and \texttt{torusLoop2PathZPow} mirror the circle’s integer powers but track each generator separately.  Crucially, the commuting relation between the two loops is encoded in the rewrite system, so integer pairs form a normal form for torus loops without the need for an additional word language.

\subsection{Loop quotients and decode/encode}

The quotient-level definitions follow the same pattern as before:

\begin{lstlisting}
abbrev TorusLoopQuot := LoopQuot Torus torusBase
abbrev torusPiOne    := PiOne   Torus torusBase

@[simp] def torusEncode : torusPiOne -> Int \times Int :=
  torusEncodeLift

@[simp] def torusDecode : Int \times Int -> torusPiOne :=
  fun z => LoopQuot.ofLoop (torusDecodePath z)
\end{lstlisting}

Here \texttt{torusEncodeLift} is the quotient lift of \texttt{torusEncodePath}.  The key lemmas establish step laws analogous to those on the circle, for example:

\begin{lstlisting}
theorem torusEncode_comp_loop1 (x : torusPiOne) :
    torusEncode (LoopQuot.comp x torusLoop1Class) =
      ((torusEncode x).1 + 1, (torusEncode x).2)

theorem torusEncode_comp_loop2 (x : torusPiOne) :
    torusEncode (LoopQuot.comp x torusLoop2Class) =
      ((torusEncode x).1, (torusEncode x).2 + 1)
\end{lstlisting}

and similarly for inverses, subtracting in the appropriate coordinate.  These statements are proved by unfolding the definitions, picking representatives, and applying the transport lemmas that describe how \texttt{torusCode} behaves along the two generators.

On the decoding side we iterate \texttt{torusLoop1} and \texttt{torusLoop2} according to the integer pair and show that this process respects addition and subtraction in each component.  The raw encode/decode pair satisfies:

\[
  \texttt{torusEncodePath (torusDecodePath z)} = z,
  \qquad
  \texttt{torusDecode (torusEncode x)} = x,
\]

with the proofs again following from normalization and the fact that the rewrite system can shuffle generator occurrences past each other.  Passing to the quotient yields the desired group isomorphism:

\begin{lstlisting}
noncomputable def torusPiOneEquivIntProd :
    SimpleEquiv torusPiOne (Int \times Int) where
  toFun := torusEncode
  invFun := torusDecode
  left_inv := torusDecode_torusEncode
  right_inv := torusEncode_torusDecode
\end{lstlisting}

Consequently \(\pi_1(T^2)\) is definitionally the product \(\mathbb{Z} \times \mathbb{Z}\) inside Lean, and the entire encode/decode algebra (addition laws, inversion, cancellation) is available as `[simp]` lemmas for subsequent developments.

\section{Fundamental group of the Klein bottle}
\label{sec:klein}

The Klein bottle is a non-orientable surface whose fundamental group is the semidirect product $\mathbb{Z} \rtimes \mathbb{Z}$. Unlike the torus, the two generators do not commute: the characteristic relation is $aba^{-1} = b^{-1}$. The Lean file \texttt{ComputationalPaths/\allowbreak Path/\allowbreak HIT/\allowbreak KleinBottle.lean}\footnote{See~\cite{ComputationalPathsLean}.} encodes this structure as an axiomatic higher-inductive type:

\begin{lstlisting}
axiom KleinBottle : Type u
axiom kleinBase : KleinBottle
axiom kleinLoopA : Path kleinBase kleinBase
axiom kleinLoopB : Path kleinBase kleinBase

axiom kleinSurf :
  Path.trans
      (Path.trans kleinLoopA kleinLoopB)
      (Path.symm kleinLoopA) =
    Path.symm kleinLoopB
\end{lstlisting}

The axiom \texttt{kleinSurf} encodes the relation $a \cdot b \cdot a^{-1} = b^{-1}$, which is the defining property of the Klein bottle as opposed to the torus. This relation implies that conjugating the vertical loop $b$ by the horizontal loop $a$ inverts it.

\subsection{Code family and transport}

The universal cover of the Klein bottle uses the same underlying set as the torus---integer pairs $\mathbb{Z} \times \mathbb{Z}$---but with different transport behavior. The code family \texttt{kleinCode : KleinBottle -> Type} is defined so that:
\begin{itemize}[nosep]
  \item Transport along \texttt{kleinLoopA} sends $(m, n)$ to $(m + 1, -n)$.
  \item Transport along \texttt{kleinLoopB} sends $(m, n)$ to $(m, n + 1)$.
\end{itemize}

The negation in the first rule reflects the relation $aba^{-1} = b^{-1}$: going around $a$ and then $b$ is equivalent to going around $b^{-1}$ and then $a$. This is captured by constructing a type equivalence that negates the second coordinate and using the univalence axiom:

\begin{lstlisting}
def kleinNegateEquiv : SimpleEquiv (Int $\times$ Int) (Int $\times$ Int) where
  toFun := fun (m, n) => (m, -n)
  invFun := fun (m, n) => (m, -n)
  left_inv := fun (m, n) => by simp
  right_inv := fun (m, n) => by simp

def kleinCodeData : KleinBottleRecData (Type _) where
  base := Int $\times$ Int
  loopA := Path.ofEq (ua (kleinShiftNegEquiv))  -- shift and negate
  loopB := Path.ofEq (ua (kleinShiftEquiv))     -- shift only
  surf := -- proof that the surface relation is satisfied
\end{lstlisting}

\subsection{Semidirect product structure}

The fundamental group $\pi_1(K)$ carries the structure of a semidirect product $\mathbb{Z} \rtimes \mathbb{Z}$. The group operation is:
\[
  (m_1, n_1) \cdot (m_2, n_2) = (m_1 + m_2, (-1)^{m_2} \cdot n_1 + n_2)
\]

This is implemented in Lean as:

\begin{lstlisting}
def kleinSemidirectMul (z1 z2 : Int $\times$ Int) : Int $\times$ Int :=
  (z1.1 + z2.1, Int.negOnePow z2.1 * z1.2 + z2.2)
\end{lstlisting}

The key theorem relating loop composition to this multiplication is:

\begin{lstlisting}
theorem kleinLoopBClass_zpow_mul_loopAClass_zpow (n m : Int) :
    LoopQuot.comp (kleinLoopBClass_zpow n) (kleinLoopAClass_zpow m) =
    LoopQuot.comp (kleinLoopAClass_zpow m)
                  (kleinLoopBClass_zpow (Int.negOnePow m * n))
\end{lstlisting}

This theorem states that $b^n \cdot a^m = a^m \cdot b^{(-1)^m \cdot n}$, which generalizes the surface relation to arbitrary integer powers.

\subsection{Encode/decode and the equivalence}

The encode map sends a loop to its integer coordinates:

\begin{lstlisting}
def kleinEncodePath :
    Path kleinBase kleinBase -> Int $\times$ Int :=
  fun p => kleinCodeToProd (kleinEncodeRaw kleinBase p)
\end{lstlisting}

The decode map constructs a loop from integer coordinates:

\begin{lstlisting}
def kleinDecodePath (z : Int $\times$ Int) :
    Path kleinBase kleinBase :=
  Path.trans (kleinLoopA_zpow z.1) (kleinLoopB_zpow z.2)
\end{lstlisting}

The two inverse laws are:
\begin{itemize}[nosep]
  \item \texttt{kleinEncode\_kleinDecodeQuot}: $\text{encode}(\text{decode}(m, n)) = (m, n)$
  \item \texttt{kleinDecodeQuot\_kleinEncode}: $\text{decode}(\text{encode}(x)) = x$
\end{itemize}

The final equivalence packages these into a single statement:

\begin{lstlisting}
noncomputable def kleinPiOneEquivIntProd :
    SimpleEquiv kleinPiOne (Int $\times$ Int) where
  toFun := kleinEncode
  invFun := kleinDecodeQuot
  left_inv := kleinDecodeQuot_kleinEncode
  right_inv := kleinEncode_kleinDecodeQuot
\end{lstlisting}

This establishes that $\pi_1(K) \cong \mathbb{Z} \rtimes \mathbb{Z}$ as sets, with the group structure determined by the semidirect product multiplication.

\section{Fundamental group of the real projective plane}
\label{sec:projective}

The real projective plane $\mathbb{RP}^2$ is obtained from a disk by identifying antipodal boundary points. Its fundamental group is the cyclic group of order two, $\mathbb{Z}_2$. The Lean file \texttt{ComputationalPaths/\allowbreak Path/\allowbreak HIT/\allowbreak ProjectivePlane.lean}\footnote{See~\cite{ComputationalPathsLean}.} models this as a higher-inductive type with a single loop whose square is trivial:

\begin{lstlisting}
axiom ProjectivePlane : Type u
axiom projectiveBase : ProjectivePlane
axiom projectiveLoop :
    Path projectiveBase projectiveBase

axiom projectiveLoopSquare :
  Path.trans projectiveLoop projectiveLoop =
    Path.refl projectiveBase
\end{lstlisting}

The axiom \texttt{projectiveLoopSquare} states that $\alpha \cdot \alpha = \text{refl}$, where $\alpha$ is the fundamental loop. This is the defining property of $\mathbb{RP}^2$: going around the loop twice returns to the identity.

\subsection{Code family over Bool}

Since the fundamental group has only two elements, the universal cover can be modeled using booleans. The code family \texttt{projectiveCode : ProjectivePlane -> Type} is defined so that:
\begin{itemize}[nosep]
  \item The fiber over the base point is \texttt{Bool}.
  \item Transport along \texttt{projectiveLoop} applies boolean negation.
\end{itemize}

The negation behavior reflects the $\mathbb{Z}_2$ structure: going around the loop once switches between the two elements of the cover, and going around twice returns to the starting point.

\begin{lstlisting}
def projectiveEquiv : SimpleEquiv Bool Bool where
  toFun := not
  invFun := not
  left_inv := Bool.not_not
  right_inv := Bool.not_not

theorem projectiveCode_transport_loop (b : Bool) :
    Path.transport projectiveLoop
        (projectiveCode_base $\blacktriangleright$ b) =
      projectiveCode_base $\blacktriangleright$ (not b)
\end{lstlisting}

\subsection{Encode and decode}

The encode map uses transport in the code family:

\begin{lstlisting}
def projectiveEncodePath :
    Path projectiveBase projectiveBase -> Bool :=
  fun p => projectiveCodeToBool
    (Path.transport p (projectiveCode_base $\blacktriangleright$ false))
\end{lstlisting}

The key computation rules are:
\begin{itemize}[nosep]
  \item \texttt{projectiveEncodePath\_refl}: Encoding the identity path yields \texttt{false}.
  \item \texttt{projectiveEncodePath\_loop}: Encoding the fundamental loop yields \texttt{true}.
\end{itemize}

The decode map interprets a boolean as either the identity or the fundamental loop:

\begin{lstlisting}
def toPathZ2 : Bool -> LoopQuot ProjectivePlane projectiveBase
  | false => LoopQuot.id
  | true  => Quot.mk _ projectiveLoop
\end{lstlisting}

\subsection{The equivalence}

The two inverse laws establish the bijection:
\begin{itemize}[nosep]
  \item \texttt{projectiveEncode\_decode}: $\text{encode}(\text{decode}(b)) = b$ for $b : \texttt{Bool}$.
  \item \texttt{projectiveDecode\_encode}: $\text{decode}(\text{encode}(x)) = x$ for loops $x$.
\end{itemize}

The first follows by case analysis on the boolean. The second uses the quotient induction principle and the fact that every loop is rewrite-equivalent to either the identity or the generator.

\begin{lstlisting}
def projectivePiOneEquivZ2 :
    SimpleEquiv (LoopQuot ProjectivePlane projectiveBase) Bool where
  toFun := projectiveEncodeQuot
  invFun := toPathZ2
  left_inv := projectiveDecode_encode
  right_inv := projectiveEncode_decode
\end{lstlisting}

This completes the proof that $\pi_1(\mathbb{RP}^2) \cong \mathbb{Z}_2$, where $\mathbb{Z}_2$ is represented as \texttt{Bool} with XOR as addition.

\section{Engineering and automation}
\label{sec:engineering}

The formalization described in this paper spans a substantial Lean code base. While the underlying mathematics is classical, encoding it in a modern proof assistant raises several engineering challenges.

\subsection{Library design}

A central design goal was to make the computational paths library usable beyond the specific case studies of the circle and the torus. To this end we separated the core path and rewrite machinery from the topological examples. The path library itself does not mention specific spaces; it only knows about types, paths, and rewrites. The circle and torus are defined in separate modules that can be imported selectively.

We also made deliberate choices about where to expose quotient types. In some parts of the library it is convenient to reason about raw paths and rewrites explicitly. In others it is more convenient to work modulo rewrite equivalence from the beginning. We expose both viewpoints and provide translation lemmas. For instance, we have lemmas that transport a function defined on raw paths to a function on equivalence classes, provided it respects the rewrite relation.

Another design consideration is universe management. The \texttt{Path} type is universe polymorphic, and the groupoid structure is formulated at an arbitrary universe level. This makes it possible to instantiate the theory both at small types such as \texttt{Circle} and \texttt{Torus} and at more complex types that may live in higher universes.

\subsection{Automation and proof search}

Several lemmas in the development involve long sequences of rewrite steps that are conceptually simple but tedious to write by hand. For these we rely on a combination of Lean tactics, bespoke rewrite tactics tailored to the computational path system, and external assistance from large language models that suggest candidate proof scripts.

We implemented small custom tactics that mimic the informal reasoning used in the original computational paths papers. One such tactic repeatedly applies the identity elimination rules and inverse cancellation rules until no further simplification is possible. Another tactic searches for opportunities to use the torus commuting lemma to swap adjacent loops and bring a word closer to normal form.

For example, a typical lemma might state that a complex composite of loops on the torus can be rewritten to a normal form. The high level structure of the proof is clear: repeatedly apply commuting and identity rewrites until the path is normalized. We encapsulate these patterns into a tactic that searches for applicable rewrite rules and applies them until no further progress is possible. This is particularly important in the torus development, where naive proofs quickly become unreadable.

\subsection{Role of AI assistance}

During the development we experimented with AI assisted proof search. Large language models can suggest candidate Lean proof scripts that are then checked by the Lean kernel. This proved particularly useful in proofs that require many small and routine steps. We emphasize, however, that all final proofs are fully checked by Lean and that the logical soundness of the development does not depend on the internal reasoning of the AI tools.

Our workflow typically proceeds as follows. For a given lemma, we write a precise statement and a rough outline of the proof in comments. We then ask an AI system to propose a Lean script that follows this outline. The proposed script rarely works unchanged, but it often provides a useful starting point that we refine by hand. We then verify the final script with Lean. In some cases the AI suggestion was substantially different from our outline but still valid, which highlighted alternative proof strategies.

This experience suggests that computational paths are a good target for AI assisted formalization: the proofs are rich enough to be mathematically interesting, yet structured enough that local search and pattern based heuristics work well.

\section{Related work}
\label{sec:related}

Our work sits at the intersection of several lines of research.

First, it contributes to the formalization of homotopy theoretic concepts in type theory based proof assistants. Formalizations of the fundamental group of the circle and torus exist in homotopy type theory and cubical type theory, often using path types as primitive~\cite{HoTTBook2013,LicataShulman2013,Cohen2018Cubical}. Our work offers an alternative based on computational paths and a rewrite based weak groupoid. It shows that the encode and decode arguments can be carried out in a setting where paths are explicit proof objects rather than primitive identity types.

Second, it complements the original theoretical work on computational paths. That line of research develops the logical system, the rewrite rules, and several applications, ranging from the explicit treatment of computational paths to their interpretation as identity types, but it does not provide a mechanization in a contemporary proof assistant~\cite{Queiroz2016Paths,Ramos2017IdentityPaths,Ramos2018ExplicitPaths,Veras2023Topological,Veras2023WeakGroupoid,Veras2019LND}. Our Lean development validates these constructions and exposes the computational details of the rewrite system. For example, the need to prove invariance of winding numbers under \texttt{RwStar} forces one to make precise which rewrite rules are allowed and how they interact.

Third, it connects to the growing body of work on AI assisted theorem proving. We use AI tools to generate candidate proofs and refactor code, while relying on Lean's kernel for final verification~\cite{deMoura2015Lean,Buzzard2020mathlib}. This workflow illustrates how large language models can assist in building substantial mathematical libraries without compromising soundness. It also suggests that computational paths, with their explicit proof terms and finite rewrite systems, may be a good playground for studying proof search strategies.

Finally, our work adds to a broader effort to formalize algebraic topology in proof assistants. There are formalizations of fundamental groups, covering spaces, and homology theories in various systems, each with its own balance between abstraction and concreteness. Computational paths occupy an interesting point in this spectrum: they are close to traditional equality reasoning, yet they admit a rich homotopical interpretation.

\section{Conclusion and future work}
\label{sec:conclusion}

We have presented a Lean 4 formalization of computational paths and their weak groupoid structure. Building on this foundation, we formalized loop spaces and fundamental groups and carried out six classical computations: the fundamental groups of the circle ($\mathbb{Z}$), the cylinder ($\mathbb{Z}$), the M\"obius band ($\mathbb{Z}$), the torus ($\mathbb{Z} \times \mathbb{Z}$), the Klein bottle ($\mathbb{Z} \rtimes \mathbb{Z}$), and the real projective plane ($\mathbb{Z}_2$). These examples span orientable and non-orientable surfaces, abelian and non-abelian groups, and infinite and finite fundamental groups. The cylinder and M\"obius band demonstrate how retraction arguments can be formalized using computational paths. The development is packaged as a reusable library that can serve as a basis for further work on computational paths and homotopy theory in proof assistants.

There are several promising directions for future work.

\begin{itemize}[nosep]
  \item \textbf{Additional spaces.} One can extend the library with further examples such as wedges of circles, higher dimensional tori, spheres of higher dimension, and more general cell complexes. This would test the scalability of the computational paths framework and its implementation. In particular, free products of groups arising from wedges of circles are a natural target, as are the higher homotopy groups $\pi_n(S^n)$.
  \item \textbf{Higher structures.} The weak groupoid structure considered here is 1 dimensional. Extending the Lean formalization to higher groupoids and higher paths would bring computational paths closer to the higher dimensional structures considered in homotopy type theory. This raises interesting questions about how to represent higher rewrite rules and coherence conditions. The surface axioms in the Klein bottle and the loop-square axiom in the projective plane are first steps toward handling 2-dimensional path algebra.
  \item \textbf{Bridging to homotopy type theory.} It would be interesting to compare the computational paths formalization with existing homotopy type theory libraries, both conceptually and through concrete translations between the two frameworks. For instance, one could attempt to construct a functor from the groupoid of computational paths to a path groupoid built from the identity type.
  \item \textbf{Improved automation.} The rewrite heavy nature of the development suggests that more sophisticated automation could be helpful. Designing tactics that understand the algebra of paths more deeply is a natural next step. For example, one could implement a tactic that recognizes words in generators and decides equality by normalizing them. The Klein bottle development, with its semidirect product structure, particularly benefits from automation that handles the sign-alternating behavior.
  \item \textbf{Integration with other libraries.} Since the formalization builds on mathlib, there is scope for integrating computational paths with other algebraic and categorical developments. For example, one could connect the fundamental groups computed here with homology computations or with higher categorical structures.
\end{itemize}

We hope that this development will serve as a starting point for a broader exploration of computational paths in formalized mathematics and that it will stimulate further interactions between proof theory, homotopy theory, and interactive theorem proving.

\backmatter

\bmhead{Acknowledgements}

We thank the Lean community for developing the tools that made this formalization possible.

\section*{Declarations}

\textbf{Funding} Not applicable.

\textbf{Conflict of interest/Competing interests} The authors declare that they have no competing interests.

\textbf{Ethics approval and consent to participate} Not applicable.

\textbf{Consent for publication} Not applicable.

\textbf{Data availability} All artifacts generated during this work are available in the \texttt{ComputationalPathsLean} repository referenced in the manuscript.

\textbf{Materials availability} Not applicable.

\textbf{Code availability} The Lean sources are publicly available at \url{https://github.com/Arthur742Ramos/ComputationalPathsLean}.

\textbf{Author contribution} All authors contributed to the conceptualization, formalization, and writing of this manuscript.

\IfFileExists{computational_paths_arxiv.bbl}{%

%
}{%
  \bibliography{references}%
}

\end{document}